# Assessing the Risks of Typhoon-Induced Multi-Hazards in South Korea

Ziyue Liu, Ph.D., Michelle T. Bensi, Ph.D.

Abstract: Tropical cyclone-induced coastal hazards can significantly damage coastal infrastructure, and these risks may intensify under future climate change. As a result, there is increasing interest in conducting comprehensive assessments of coastal hazards—including storm surge, storm wind, storm rainfall, and their combined impacts—associated with tropical cyclone events. Risk assessments that overlook the compounding nature of these hazards may lead to ineffective or insufficient mitigation strategies. This study seeks to identify and evaluate the available data, models, and methodologies for assessing both individual and compound typhoon-induced hazards in South Korea. Particular effort is devoted to exploring how established approaches from the North Atlantic region can be adapted, integrated, and extended for application in the South Korean context. Multiple sites across South Korea are analyzed to illustrate the strengths and limitations of these methods.

## 1 Introduction

Infrastructure in South Korea near coastal regions can be exposed to typhoon- (or, more generally, tropical cyclone (TC)) induced hazards. Recently, several studies have focused on predicting the behavior of typhoons in the Western North Pacific (WNP) region. Choi et al. (2019) predicted that while the total number of TC landfalls in the East Asian coastal area will decrease, the number of high-intensity TC landfalls will increase under climate change. This conclusion mirrors recent findings in the North Atlantic region (Knutson et al. 2013; Xi et al. 2024) and more generally (Knutson et al. 2020), particularly with regard to the intensity of storms. Chen et al. (2021) utilized a statistical dynamic synthetic TC model and various global climate models to perform predictive analysis of typhoon behavior under the scenario of the least constrained emission. They found that the proportion of high-intensity typhoons can be increased by up to 31% by the end of this century. Magee et al. (2021) applied multivariable Poisson regression to predict typhoon frequency and found that a flexible and adaptive modeling framework is important in typhoon prediction in different sub-regions in the WNP region. In addition, several studies have focused on creating synthetic datasets for the WNP region to support typhoon hazard assessments and overcome the limitations of the scarcity of historical typhoon events data (Bloemendaal et al. 2020; Kim and Suh 2018; Ng and Leckebusch 2021). For typhoon-induced hazards, Kim and Suh (2016) applied a random synthetic TC generation scheme to assess storm surge vulnerability in South Korea.

TC-induced Coastal hazards are inherently compound events, meaning TCs cause a combination of hazards such as locally intense precipitation, storm surges, and river flow. This complexity has led to a growing need for comprehensive assessments of coastal hazards resulting from typhoon events in the South Korean region and throughout the world. This study aims to leverage and identify opportunities to adapt and extend existing probabilistic analysis methods of TC-induced hazards to the South Korean region, building on the foundation of established methodology in the North Atlantic region.

The predominant method for assessing the frequency of TC-induced coastal hazards (particularly storm surge) in the North Atlantic has been the Joint Probability Method (JPM) (Myers 1954; Ho and Myers 1975; Russell 1969; Myers 1970). The JPM is a probabilistic modeling approach that accounts for the frequency of TC events, the joint probability distribution of TC parameters, and the distribution of coastal hazard response quantities, which can be obtained using numerical, empirical, or other models. Under the JPM, annual exceedance frequencies are calculated by applying the Theorem of Total Probability. Since its inception, improvements have been made to the JPM to enhance its efficiency and accuracy (e.g., Resio 2007; Toro 2008; Toro et al. 2010). Recently, a series of studies (Liu et al. 2024a; b, 2025a; b; Nadal-Caraballo et al. 2015, 2019, 2020, 2022b; a) have leveraged multiple statistical and machine-learning techniques to advance the state-of-the-art in this area, including expanding the JPM to assess compound coastal hazards.

In parallel, a separate subset of studies has focused on the assessment of compound hazards using modeling approaches that perform statistical assessments directly on coastal response quantities (e.g., ocean-side water level (O-sWL) or precipitation depths) (Juma et al. 2021; Lawrence 2020). These analyses are typically based on Extreme Value Analysis (EVA) methods. Unlike the JPM, these methods do not explicitly account for the mechanistic processes involved in the generation of coastal hazards; however, they are generally much less resource-intensive to apply. Extension of these methods to assess compound hazards has generally involved the application of multivariate distributions or copula-based approaches, including modeling augmented by heuristics (Bender et al. 2016; Hurk et al. 2015; Jane et al. 2020; Wadey et al. 2015).

This study explores the opportunities for applying existing TC-induced compound coastal hazard methods to the South Korean region by considering data availability and characteristics as well as the applicability and potential challenges associated with available methods.

## 2 Data Collection

In this study, efforts have been made to explore data sources in the South Korean region to support typhoon hazard assessments using JPM or direct statistical analyses. Given the compound nature of typhoon events, the data explored includes historical TC datasets, rainfall data, ocean surface water level (O-sWL) data, and wind data. Additional analysis has been performed to characterize and organize collected data.

### 2.1 Data sources

Multiple data sources have been identified for the South Korean region. Table 1 lists the data sources explored and collected in this study. It is noted that only recent records from the Korea Meteorological Administration (KMA), Korea Hydrographic and Oceanographic Agency (KHOA), and Water Management Information System (WAMIS) are publicly available. To obtain more complete datasets, a request can be sent directly to agencies for research purposes. In this study, selected requests were made to obtain additional data.

**Table 1: Identified data resources**

| Category | Source | Accessible Format |
|---|---|---|
| Historical TC tracks | International Best Track Archive for Climate Stewardship (IBTrACS) https://www.ncei.noaa.gov/products/international-best-track-archive | TXT file |
| | KMA typhoon https://apihub.kma.go.kr/ | API |
| Synthetic TC tracks | STORM algorithm generated tracks https://doi.org/10.4121/uuid:82c1dc0d-5485-43d8-901a-ce7f26cda35d | TXT file |
| O-sWL | KHOA http://www.khoa.go.kr/oceangrid/khoa/takepart/openapi/openApiObsSurveyDataInfo.do | API/TXT file |
| Local wind speed | KHOA http://www.khoa.go.kr/oceangrid/khoa/takepart/openapi/openApiObsSurveyDataInfo.do | API/TXT file |
| Rainfall | KMA Ground observation https://apihub.kma.go.kr/ | API |
| River | WMIS watershed precipitation daily http://www.wamis.go.kr:8080/wamisweb/rf/w4.do http://wamis.go.kr/ENG/RF_DUBRFOBS_PAST.do | API/web tabulated |
| | WAMIS river water level http://www.wamis.go.kr:8080/wamisweb/rf/w9.do http://wamis.go.kr/ENG/WL_DUBWLOBS_PAST.do | API/web tabulated |
| | WAMIS river flow http://www.wamis.go.kr:8080/wamisweb/rf/w15.do http://wamis.go.kr/wkw/wkw_flwsrrs_lst.do | API/web tabulated |

## 2.2 Comparison of typhoon data from multiple sources

Multiple sources of historical TC data are available in the WNP region. These data sources provide information regarding historical TC tracks and parameters. The IBTrACS project (Demuth et al. 2006; NOAA 2021) stands as the most comprehensive worldwide repository of TC information. It consolidates historical data from various agencies, forming a cohesive and publicly accessible dataset known as the "best-track dataset." This dataset enhanced the comparability between different agencies and was collaboratively developed with the World Meteorological Organization (WMO) Regional Specialized Meteorological Centres and various global organizations and individuals.

In parallel, KMA conducts forecasting and real-time analysis of TCs using satellite data, radar analysis, and both subjective and objective Dvorak techniques (Lee et al. 2019). KMA provides information (typically at a 3 to 6 hour-time step) on TCs that have occurred since 1977 (KMA 2023), including observation time, longitude, latitude, center pressure ($P_c$), maximum wind speed ($V_w$), wind radius, and direction (Jung et al. 2019). Additionally, KMA provides best-track data from 2015 based on the most current observations and methods (Kim et al. 2022). It is noted that the KMA best-track data has also been added to the IBtrACS dataset in 2024.

It is noted that differences exist among data published by different agencies in the WNP, which is caused by variations in measurement technology and recording frequency (NOAA 2021). To visualize these differences, an exploratory analysis of temporal coverage among multiple sources of storm data is conducted. Figure 1 shows a comparison of the fraction of (a) $V_w$ and (b) $P_c$ observations associated with a storm that are recorded in datasets from multiple agencies providing information for West-Pacific region storms. The IBTrACS dataset is the source for the following data: WMO; USA (which includes data provided by US National Hurricane Center, Joint Typhoon Warning Center (JTWC), Central Pacific Hurricane Center and other USA agencies); TOKYO (WMO Regional Specialised Meteorological Center at Tokyo); CMA (Chinese Meteorological Administration Shanghai Typhoon Institute); and HKO (Hong Kong Observatory). KMA data are downloaded from KMA's website. It is noted that the open-access portion of the data in the KMA API is only available after 2001.

The historical storm observation fraction plots (Figure 1) reveal that the USA data source has the longest record history of wind velocity ($V_w$) dating back to 1945, while CMA holds the longest record history of central pressure ($P_c$) (from 1949 to present). However, the CMA data lacks storm size data, which is available from other sources.

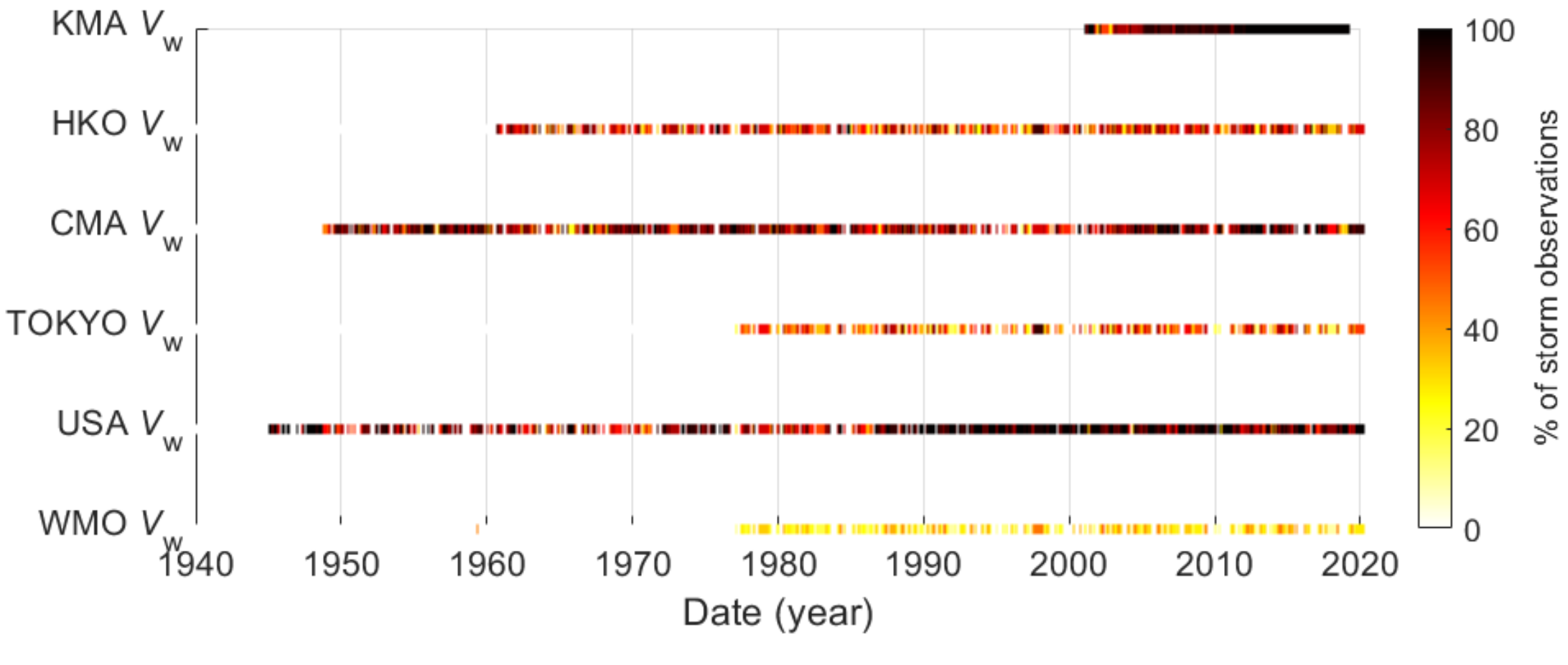

(a)

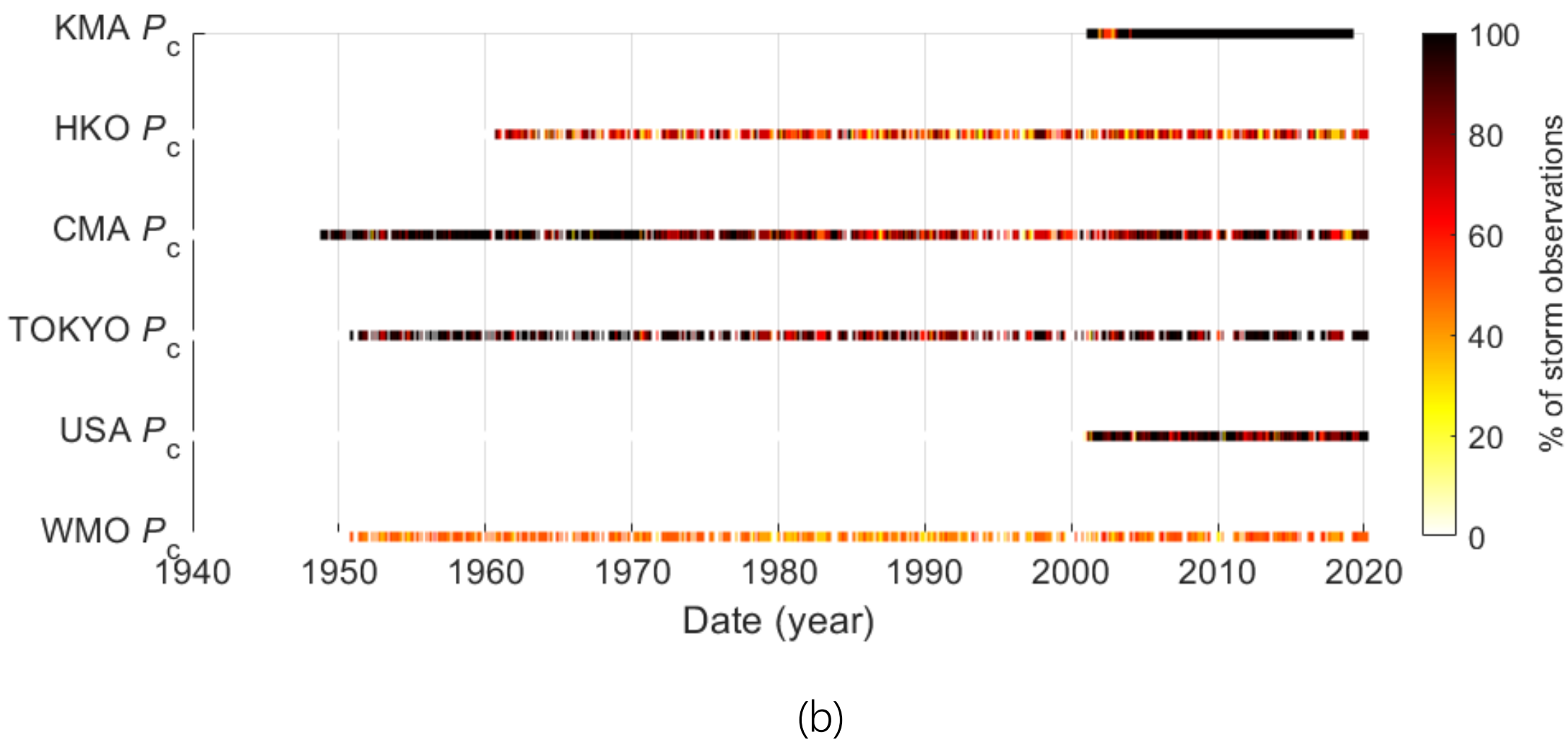

**Figure 1. Temporal historical storm observation coverage, expressed as ratios of track points that include (a) $V_{\mathrm{w}}$ and (b) $P_{\mathrm{c}}$ for each storm in the available datasets. Visualization strategy based on Liu et al. (2024b).**

To further explore the differences in recorded values of storm parameters across datasets published by various agencies, we plot the time-series of $V_{\mathrm{w}}$ and $P_{\mathrm{c}}$ from multiple agencies for selected storms for comparison. The time-series of $V_{\mathrm{w}}$ (for selected storms) is presented in Figure 2, while the time-series of $P_{\mathrm{c}}$ is presented in Figure 3.

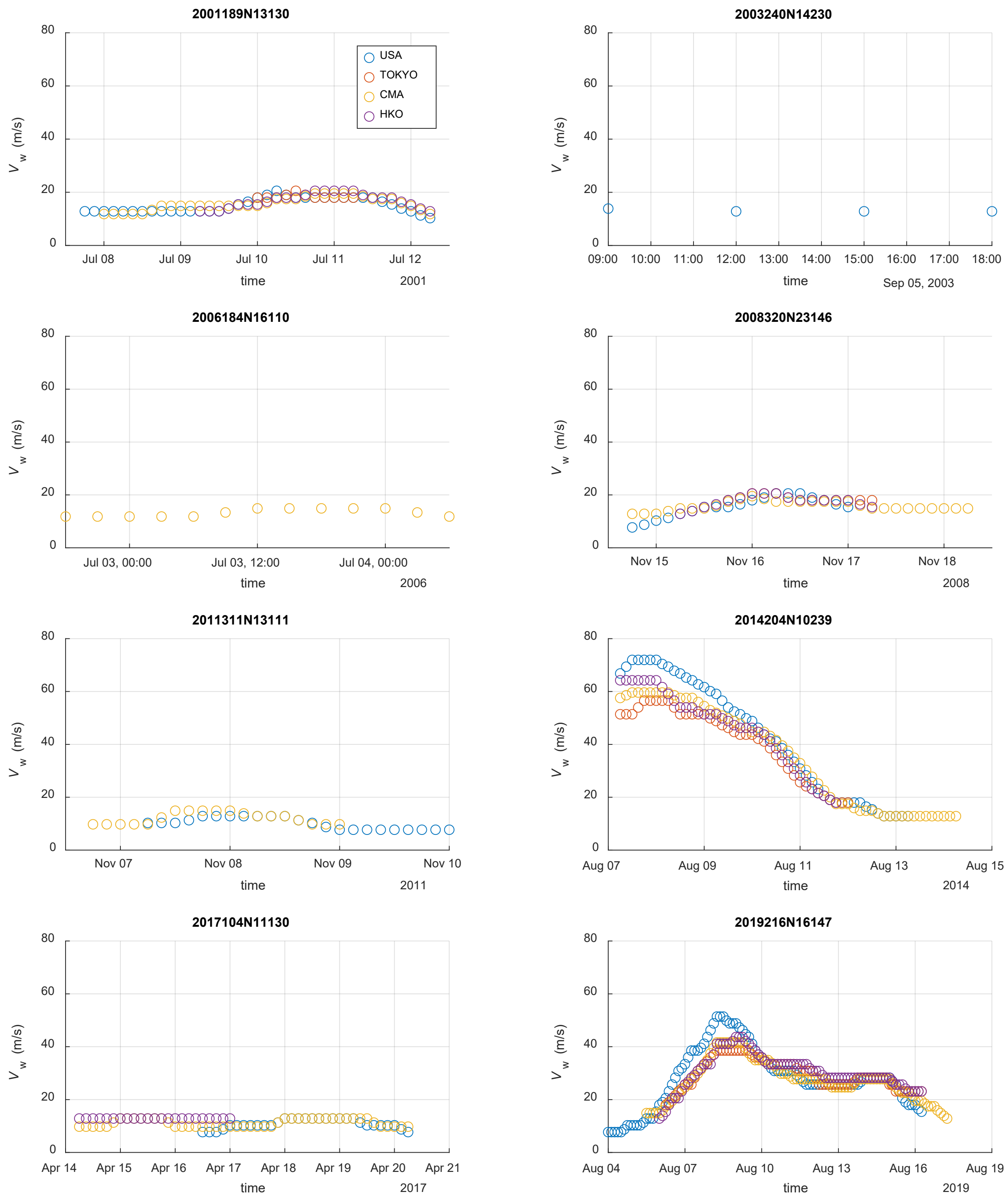

** NOTE: Windspeed metrics published by agencies use different windspeed averaging intervals. US agencies publish 1-minute values. CMA publishes 2-minute values. TOKYO (JMA) publishes 10-minute values.*

**Figure 2. $V_w$ time-series of selected storms from multiple agencies. The title of each subplot is a storm identifier assigned by the IBTrACS algorithm.**

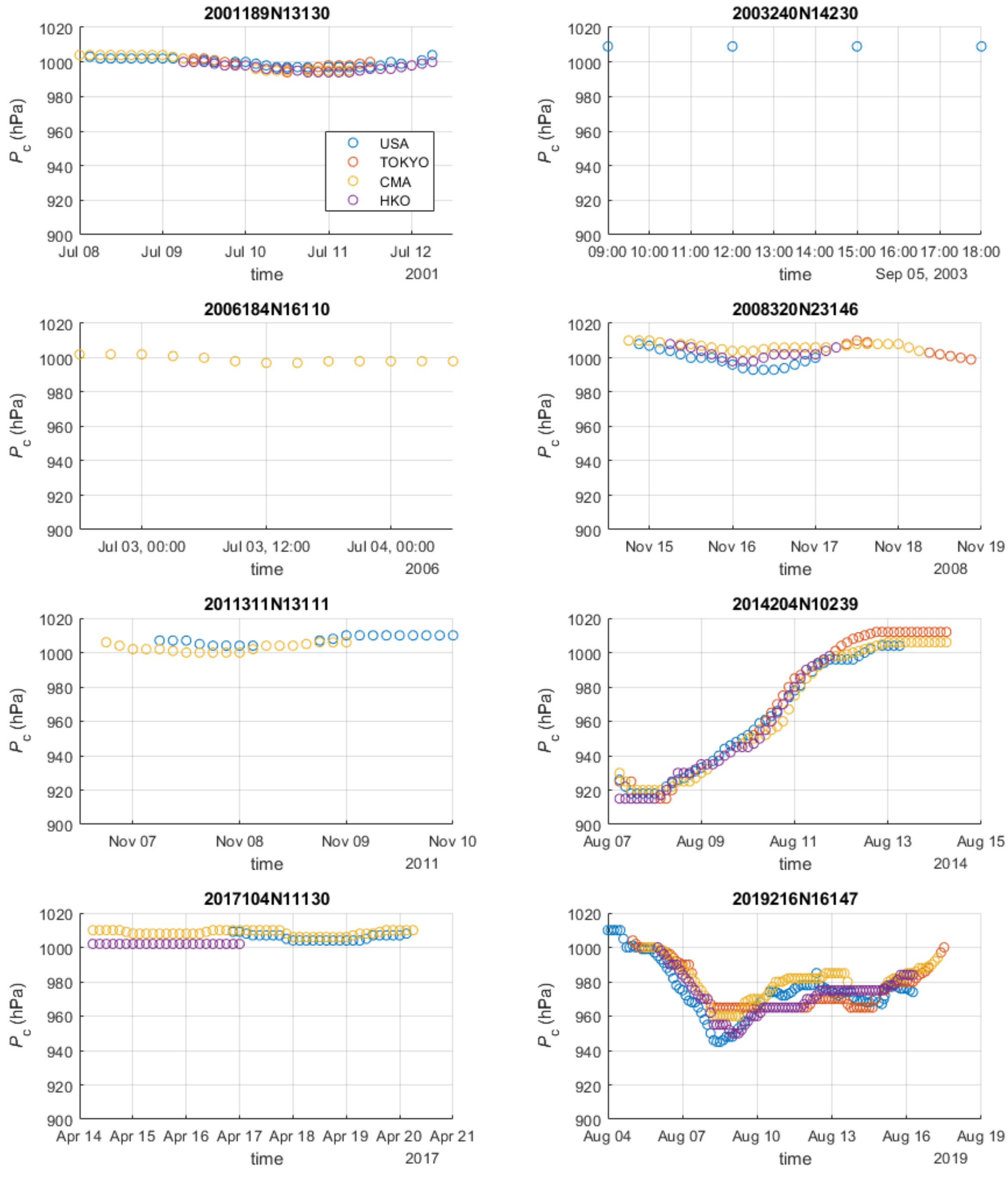

**Figure 3. $P_c$ time-series of selected storms from multiple agencies. The title of each subplot is a storm identifier assigned by the IBTrACS algorithm.**

The $V_w$ time-series presented in Figure 2 suggest that the USA-sourced values tend to assign a higher $V_w$ value than other agencies, due to the particular metrics that are reported. To illustrate this point further, a scatter plot comparing the USA- to TOKYO-sourced data (representing WMO in this region) is displayed in Figure 4.

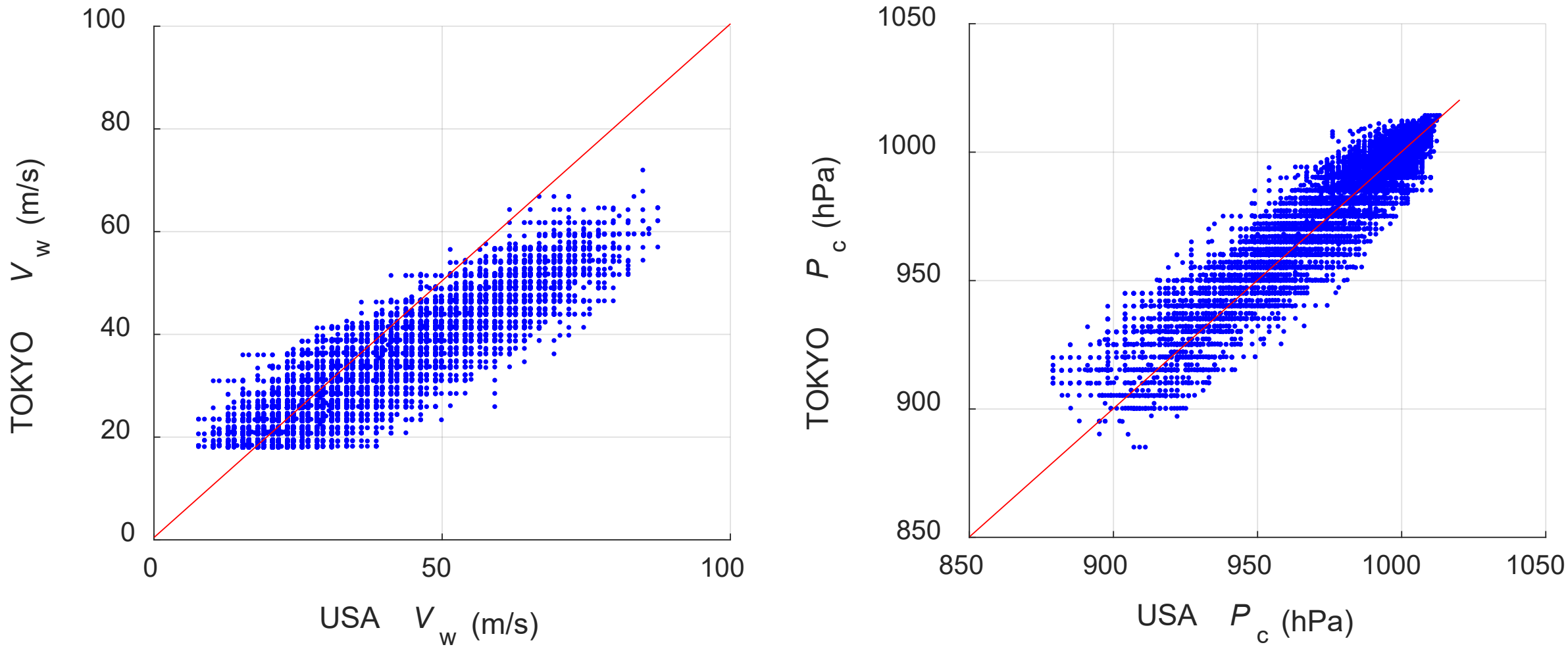

*NOTE: USA $V_w$ values reflect 1-minute average wind speeds. TOKYO (JMA) $V_w$ values reflect 10-minute average wind speeds.*

(a) (b)

**Figure 4. Scatter plot of (a) $V_{\mathrm{w}}$ data and (b) $P_{\mathrm{c}}$ data, comparing the data sources from USA and TOKYO agencies. All available data in the West-Pacific region is included.**

It can be observed that the USA-sourced $V_{\mathrm{w}}$ values are generally greater than TOKYO-sourced $V_{\mathrm{w}}$ values, while this discrepancy is not observed for $P_{\mathrm{c}}$. This discrepancy regarding $V_{\mathrm{w}}$ data is judged to be attributed to the averaging approach employed by the USA agency (NOAA and JTWC), which records $V_{\mathrm{w}}$ as a one-minute average, while other agencies use longer averaging periods (e.g., TOKYO utilizes a 10-minute average). The difference in averaging periods underscores the need for careful consideration when comparing $V_{\mathrm{w}}$ values and other data series across agencies.

### 2.3 TC track data imputation in West-Pacific basin

Historical storm datasets exhibit incomplete records due to short recording durations, low spatial resolution in observations, and missing data for certain historical storms. This can affect hazard assessment results. For example, essential storm parameters for JPM analysis (i.e., $P_{\mathrm{c}}$ and $R_{\max}$) are frequently unavailable for large portions of the historical storm dataset. The absence of these observations can adversely impact the outcomes of statistical evaluations of storm parameters, potentially introducing bias into the resulting analyses. This challenge is exacerbated when constructing joint probability models aiming to comprehensively capture the complete dependence among random variables, as highlighted by Liu et al. (2024a).

Liu et al. (2024b) developed a machine learning-based data imputation method to address the aforementioned data limitations, building upon the framework established by Nadal-Caraballo et al. (2020, 2022b). More recently, Agrawal et al. (2025) applied different deep learning approaches to impute missing values of $R_{\max}$ at the time series scale.

In this study, a machine-learning based data imputation was applied to the West-Pacific region using USA agency-published data to enrich the TC statistical sampling data in study region. Figure 5 shows the effect of this data imputation process on statistical sampling data, which is generated based on a distance adjustment process developed by the United States Army Corps of Engineers (USACE) Coastal Hazard System-Probabilistic Coastal Hazard Analysis (CHS-PCHA) (Nadal-Caraballo et al. 2015).

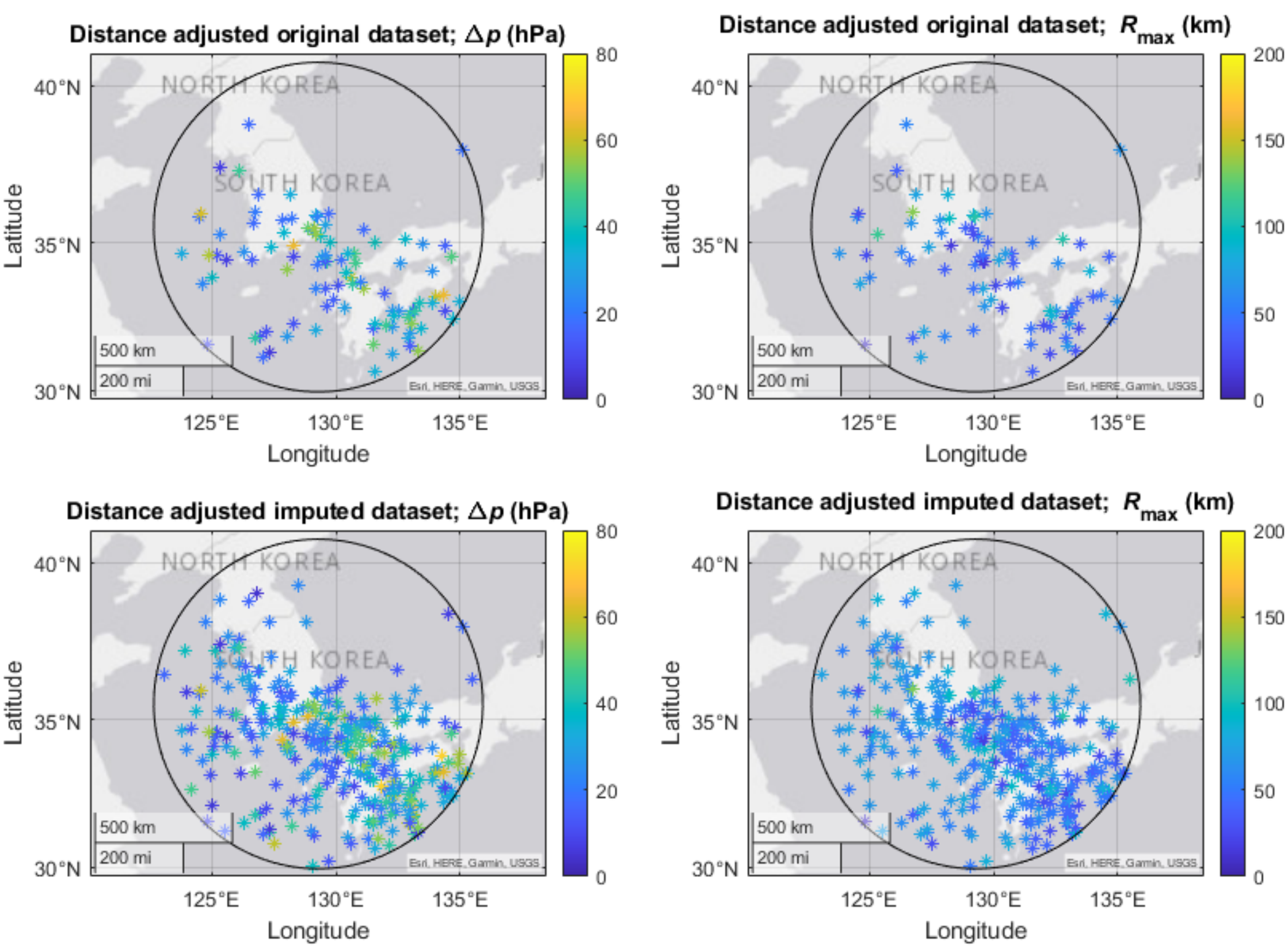

**Figure 5. Data imputation effect on statistical sampling data. The scatter plot maps show the statistical sampling data locations (lat/lon) and values (color bar) for $\Delta p$ and $R_{\max}$ using the original dataset and imputed dataset.**

## 3 Analysis

In this section, a series of statistical analyses needed for JPM inputs and direct statistical assessments of historical observation (EVA) were performed to identify potential opportunities and limitations for leveraging these approaches in the South Korean region.

### 3.1 JPM-based TC probabilistic analysis

JPM is a probabilistic modeling approach that applies the Theorem of Total Probability to estimate frequencies of exceedance by convolving the storm recurrence of TC events, the joint probability distribution of TC parameters, and the probability distribution of coastal hazards response quantities obtained using numerical, empirical, or other models.

#### 3.1.1 TC storm recurrence rate

To directly visualize typhoon occurrence frequency in the study region (as is needed for a JPM analysis), storm recurrence rates (SRR) with a 200 km radius capture zone (noted as $SRR_{200km}$) are computed based on the method of Nadal-Caraballo et al. (2019), which has been commonly applied in the North Atlantic region. Results are presented as a grided map and plotted in Figure 6 with multiple nuclear power plant (NPP) sites marked as reference location of critical infrastructures. It can be observed that the southeast coast of South Korea is under a higher SRR than other coastal regions of South Korea. Further analyses can be performed to explore the sensitivity of results to the modeling approach used (e.g., capture zone versus kernel-based approaches) and associated assumptions (e.g., optimal kernel bandwidth) to understand differences between the North Atlantic and WNP. However, the utility of these further studies depends on the practicality of employing the JPM approach, given overall data and model availability in the region.

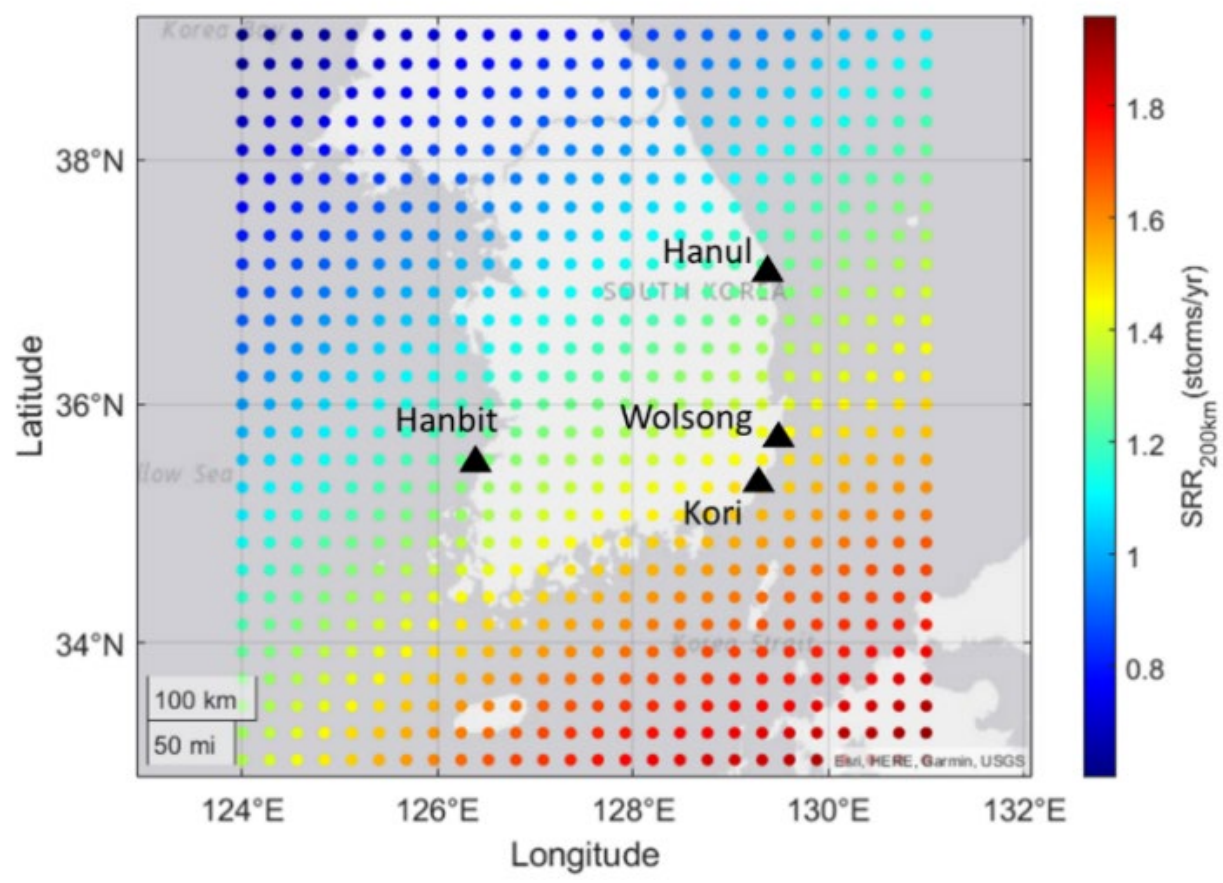

**Figure 6. SRR map in the region near South Korea; Black triangles mark sites of NPPs in South Korea.**

### 3.1.2 TC parameter joint distributions

Leveraging the imputed historical storm datasets, a statistical analysis has been performed for the South Korean region using established methods (Nadal-Caraballo et al. 2020, 2022b). The site of the Kori NPP at Busan ([35.33, 129.29]) is used as the coastal reference location (CRL) for an illustrative application. A Gaussian kernel function-based distance weight adjustment method (Chouinard and Liu 1997; Nadal-Caraballo et al. 2015) is used to generate geographically-weighted statistical sampling data. This statistical analysis considers data within 600 km of the CRL, as presented in Figure 7.

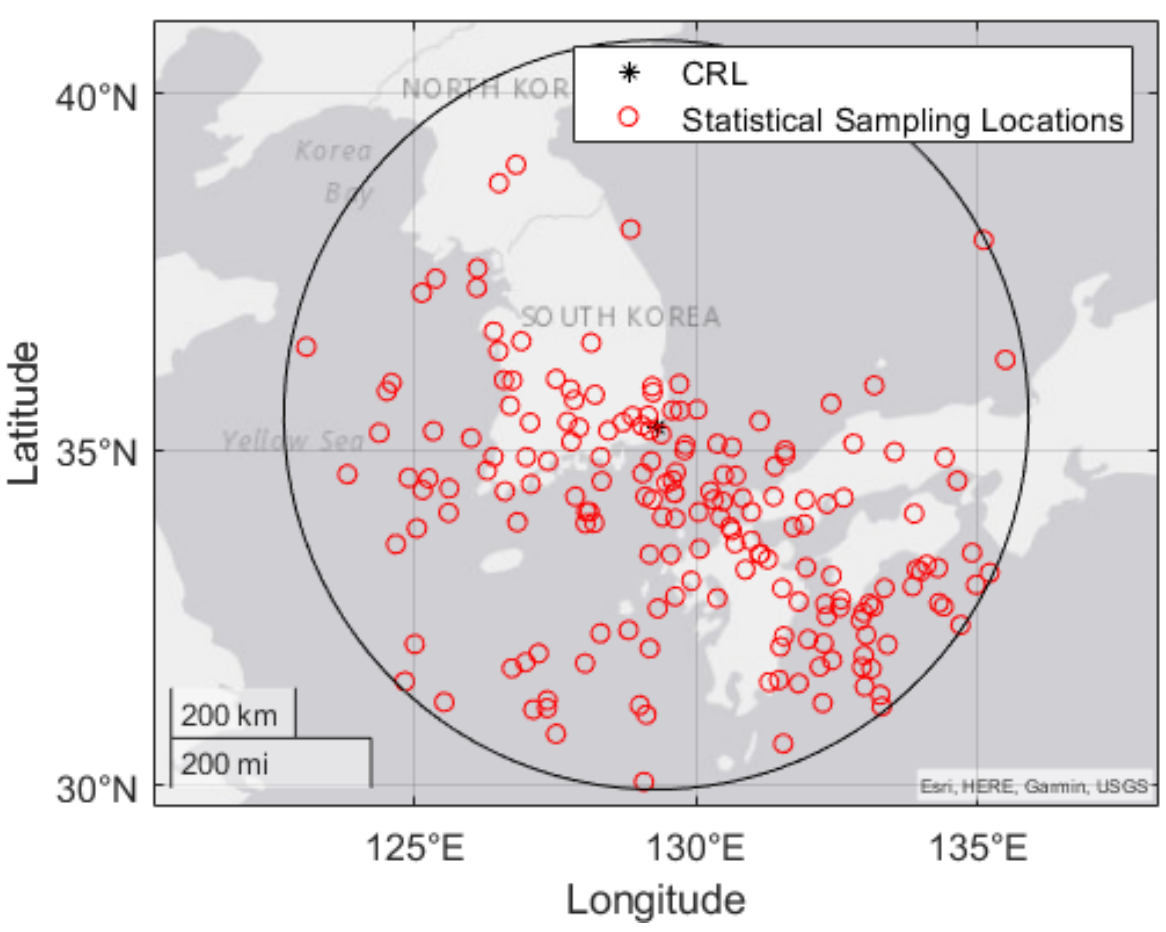

**Figure 7. Center of statistical analysis of the study region (i.e., the CRL), locations of sample data, and capture zone (the black circle) of the CRL.**

Marginal distributions have been fitted for the storm parameters central pressure deficit ($\Delta P$), $V_{\mathrm{f}}$, $V_{\mathrm{w}}$ and $\Theta$ using distribution assumptions informed by analyses performed for the North Atlantic. The truncated Weibull distribution is used for $\Delta P$ because of its flexibility in modeling the heavy tail of $\Delta P$. The log-normal distribution is used for $V_{\mathrm{f}}$ and $R_{\max}$. For heading direction $\Theta$, a directional storm recurrence rate (DSRR) (Chouinard and Liu 1997) is used to represent its probability model as:

$$\lambda_{\theta} = \frac{1}{T}\sum_{i}^{n} w(d_i)\, w(\theta_i - \theta) \tag{1}$$

where $\lambda_{\theta}$ is the DSRR of heading direction $\Theta$, $T$ is the record length (yr) of historical data, $w(d_i)$ is a Gaussian kernel function distance-based weight, and $d_i$ is the distance between the location associated with the data point $i$ and the CRL (Nadal-Caraballo et al. 2019). Leveraging the work of (Liu et al. (2024a), a von Mises Kernel function (Taylor 2008) is used for $w(\theta_i - \theta)$. Figure 8 shows the fitted marginals, which suggests that the specified distributions generally perform well. However, some deviations between the empirical and fitted cumulative distribution functions (CDFs) suggest potential opportunities for alternate assumptions or refinements.

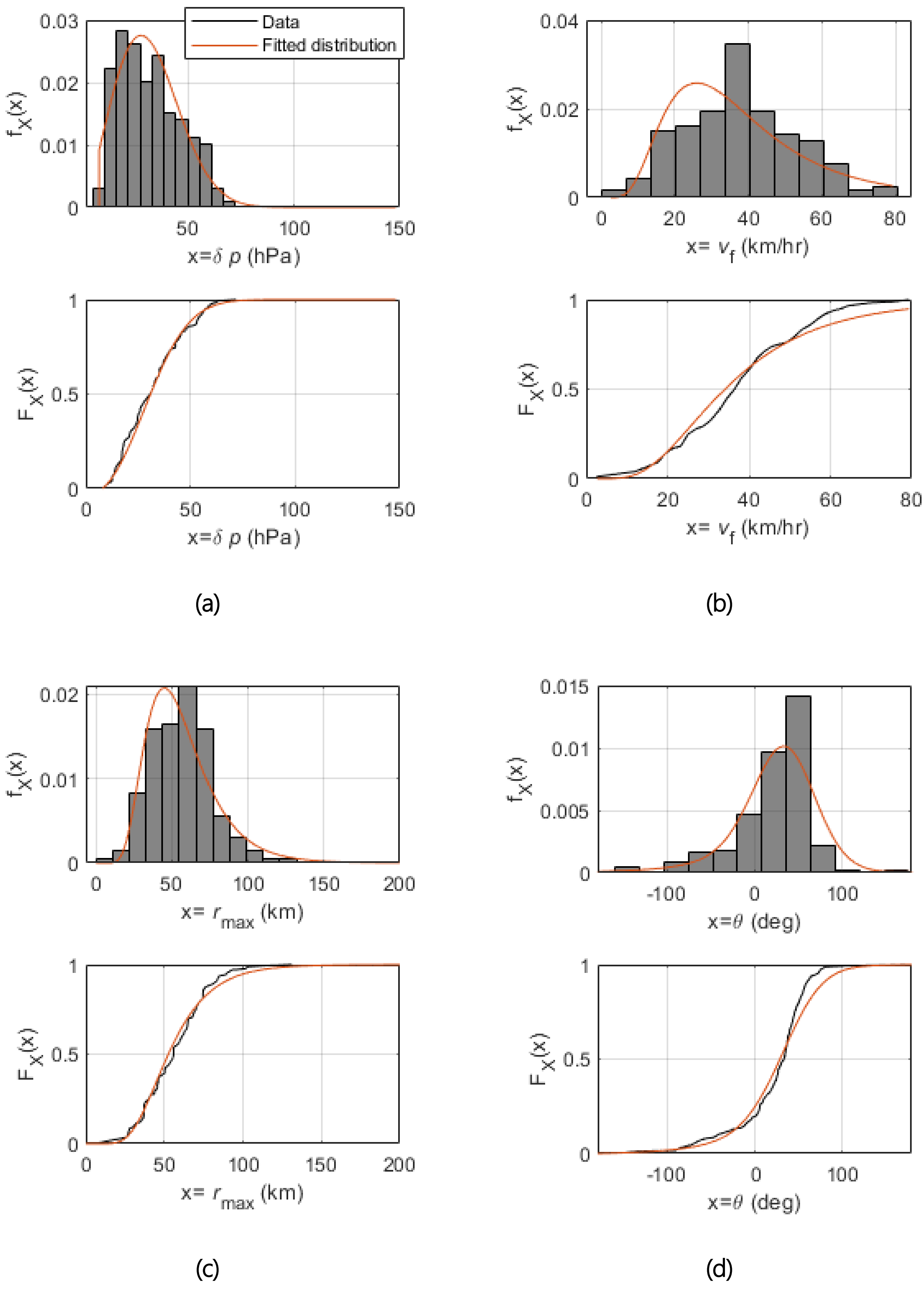

**Figure 8. Marginal distributions fitted for storm parameters (a) $\Delta P$; (b) $V_{\mathrm{f}}$; (c) $R_{\mathrm{max}}$; (d) $\Theta$.**

Considering that the data imputation process might enhance the correlation between storm parameters and reduce data variability (Liu et al. 2024c), this study has employed marginal distributions generated from the imputed dataset. However, the correlation estimated using the original dataset is used for joint probability analysis. Table 2 lists the estimated correlation using the original dataset and imputed dataset (shown in brackets).

**Table 2. Estimated correlation (Kendall's $\tau$) from sampling data.**

| | $\Delta P$ | $V_{\mathrm{f}}$ | $R_{\max}$ | $\Theta$ |
|---|---|---|---|---|
| $\Delta P$ | 1 | 0.17 [0.22] | -0.32 [-0.47] | 0.00 [-0.01] |
| $V_{\mathrm{f}}$ | 0.17 [0.22] | 1 | -0.03 [-0.17] | 0.18 [0.22] |
| $R_{\max}$ | -0.32 [-0.47] | -0.03 [-0.17] | 1 | -0.04 [-0.06] |
| $\Theta$ | 0.00 [-0.01] | 0.18 [0.22] | -0.04 [-0.06] | 1 |

*Note: Values inside the bracket are computed using the imputed dataset.*

Table 2 shows that the data imputation process enhances the correlation between storm parameters to different extents. The influence on correlations unrelated to $\Delta P$ and $R_{\max}$ can be attributed to the alteration in the data sampling locations due to data imputation. This change affects the sampled data of storm parameters other than $\Delta P$ and $R_{\max}$.

Leveraging methods employed in recent JPM studies of storm parameters (Liu et al. 2024b; Nadal-Caraballo et al. 2020, 2022b), a copula analysis is conducted for TC parameters in this case study. A copula can be regarded as a function that couples multiple marginal distribution functions to create a multivariate distribution function. According to Sklar's theorem (Sklar 1959), the joint cumulative distribution function (CDF) $H(\square)$ can be expressed as a closed form function as:

$$H(x_1,\ldots,x_{\mathrm{n}}) = C(F_1(x_1),\ldots,F_{\mathrm{n}}(x_{\mathrm{n}})) \quad (2)$$

where: $C(\square)$ is a copula function; $x_i$, $i = 1,\ldots,n$ are realizations of random variables $X_i$; $F_i(x_i)$ is the CDF of $X_i$.

In this work, a meta-Gaussian copula (MGC) is employed as the primary copula model for TC parameters. For comparative analysis, a linear-circular Gaussian vine copula (LCGV) is fitted using the method described in (Liu et al. 2024b). Within this LCGV, a quadratic section linear-circular copula is utilized to characterize potential linear-circular dependences between $\Theta$ and other linear storm parameters. The log-likelihood of both copula models is computed for comparison. The log-likelihood of the $j$th copula model is calculated as:

$$\mathcal{L}^j = \sum_{i=1}^{n} \ln(f_i^j) \quad (3)$$

where $f_i^j$ is the Probability Density Functions (PDF) value of the $j$th copula model at the $i$th statistical sampling points, $n$ is the number of statistical sampling points. In Figure 9 and Figure 10, the pair-wise joint PDF contours are plotted for both copula models with corresponding log-likelihood.

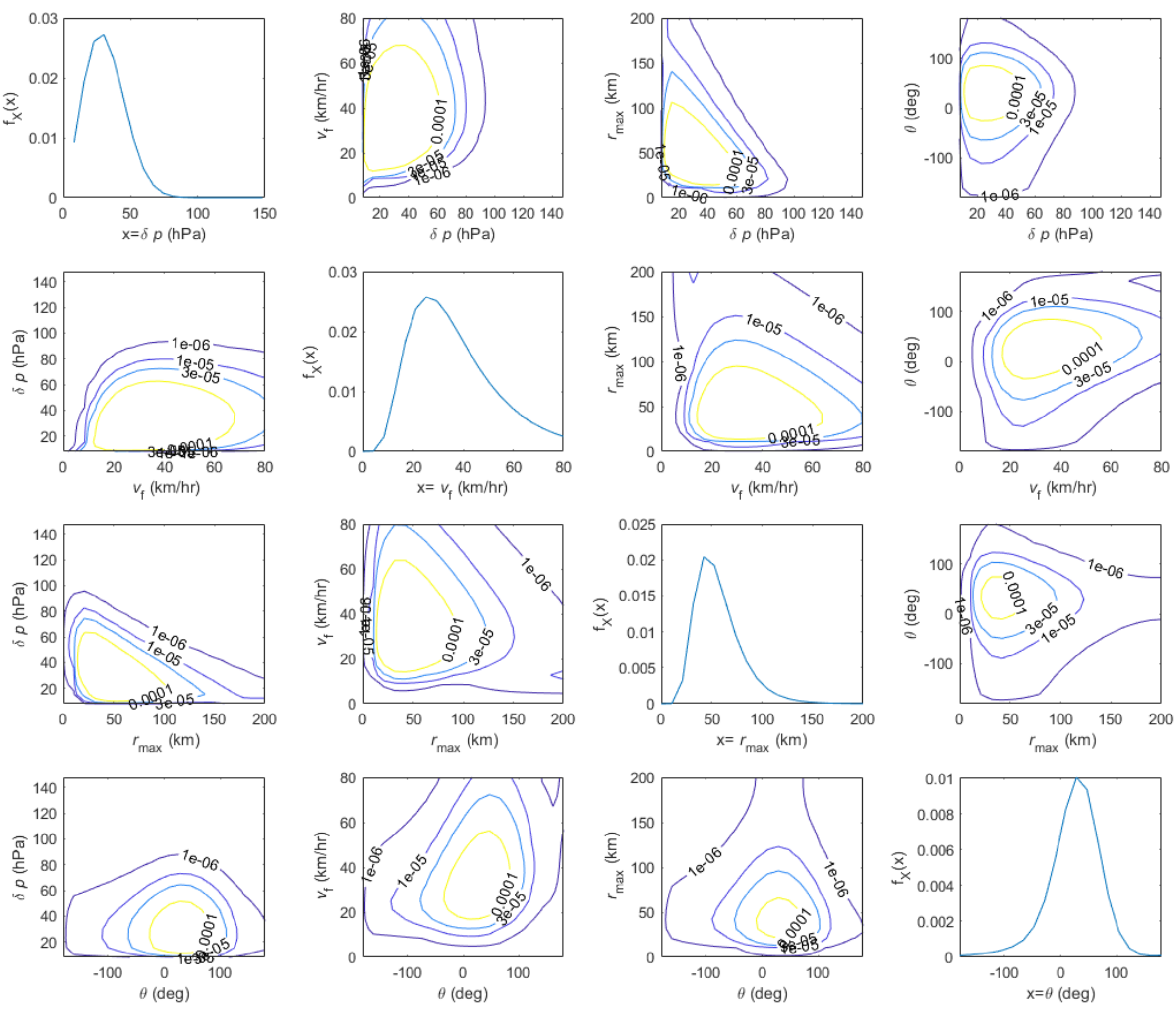

**Figure 9. MGC model pair-wise joint PDF contour; $\mathcal{L}$ =-1.42e3.**

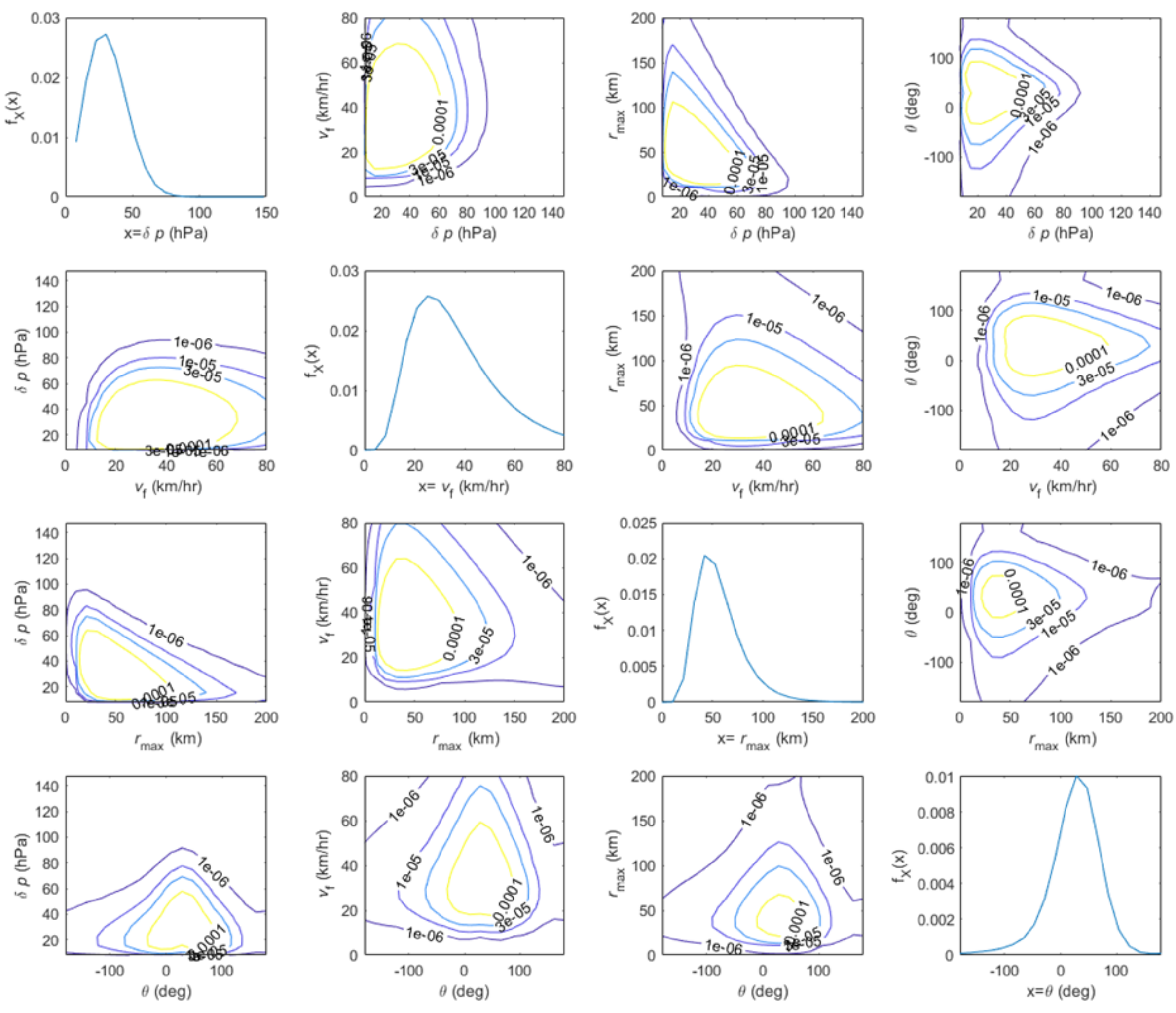

**Figure 10. LCGV model pair-wise joint PDF contour; $\mathcal{L}$ =-1.41e3.**

### 3.2 TC-induced multi-hazards analysis

In this study, local observational records are utilized to conduct a multi-hazard joint probability analysis for TC-induced surge, wind, and rainfall. First, the tidal effects in the study region are investigated. Subsequently, a copula-based approach is applied to analyze the compound behavior of the three variables. Figure 11 shows the location of city sites involved in this analysis.

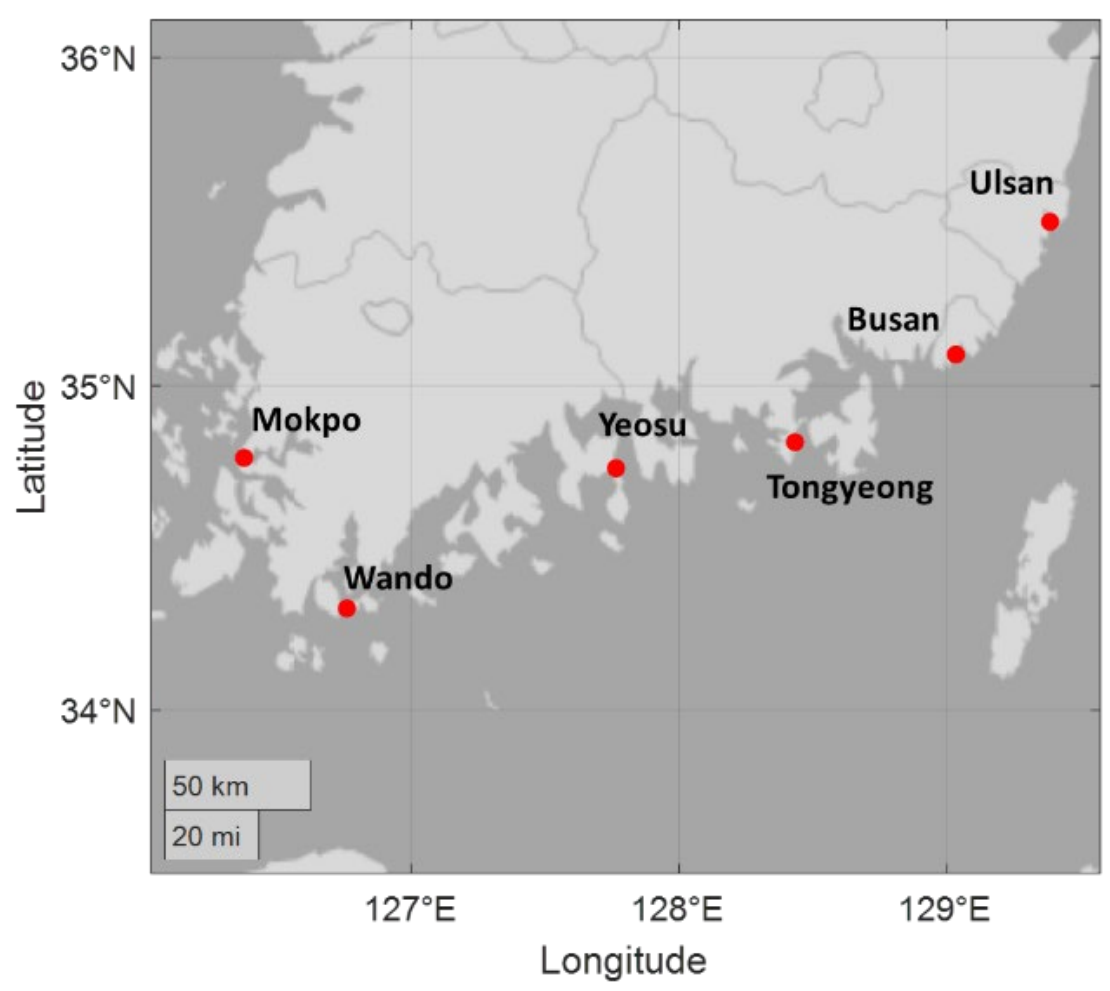

**Figure 11. Gauge location.**

### 3.2.1 Tidal effect

The O-sWL data plays an important role in this multi-hazard analysis, as it is used as source of TC surge. It is noted that O-sWL data in South Korea generally exhibit a wider range than those in well-studied North Atlantic regions like the Gulf region or Florida (e.g., Jane et al. 2020; Vaidya et al. 2023) primarily due to stronger astronomical tides. These strong tidal effects can be attributed to South Korea's higher latitude and complex coastal bathymetry, especially near locations such as Mokpo, Yeosu, Wando, and Tongyeong. As a result, it is useful to distinguish between total water levels and storm surge components. Figure 11 present examples of predicted and observed O-sWL data over six months, showing strong diurnal and semidiurnal tides, along with a semimonthly variation linked to lunar cycles (NOAA 2024; Pugh 1987).

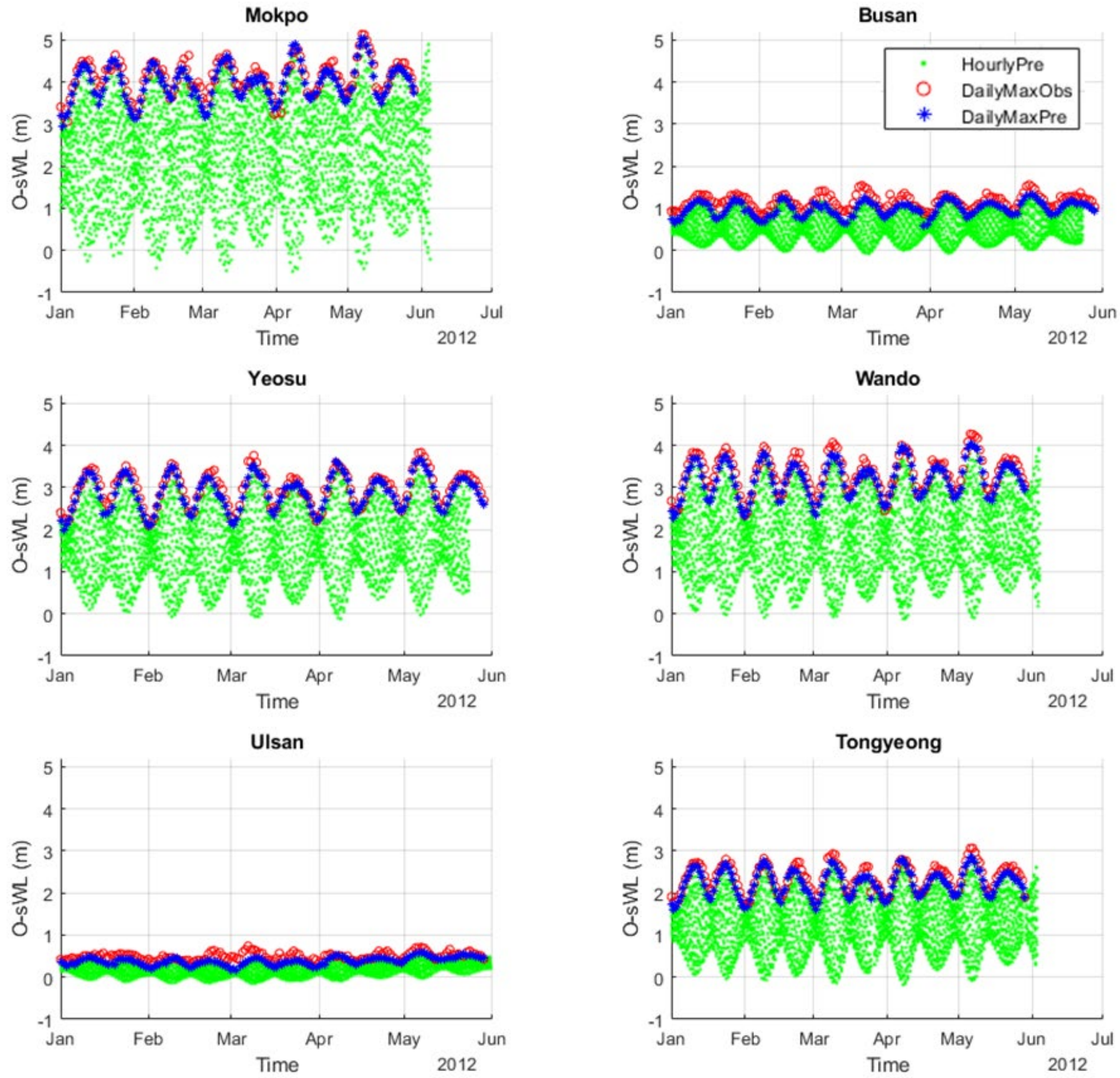

**Figure 12. Astronomical tidal effect on O-sWL data.**

It is important to note that semimonthly tidal variations introduce substantial noise when directly calculating the correlation between observed sea water level (O-sWL) and other variables (e.g., wind speed and rainfall). Therefore, non-tidal residuals (NTRs) is used for analysis. In this work, NTRs are obtained by subtracting the predicted tidal component from the observed O-sWL and represents the portion of O-sWL that effectively reflects the storm surge component (Pugh 1987). Because the predicted tidal component data (which is needed for NTRs calculation) from KHOA are only available after 2012, this study uses local observational records from 2012 onward for the multi-hazard analysis.

### 3.2.2 Trivariate analysis

In this study, a trivariate analysis considering TC surge, TC wind and TC rainfall is performed. A latitude–longitude bounding box is defined to extract storm events near the study region using the IBTrACS dataset. Figure 13 shows the map of the defined bounding box and extracted storm events.

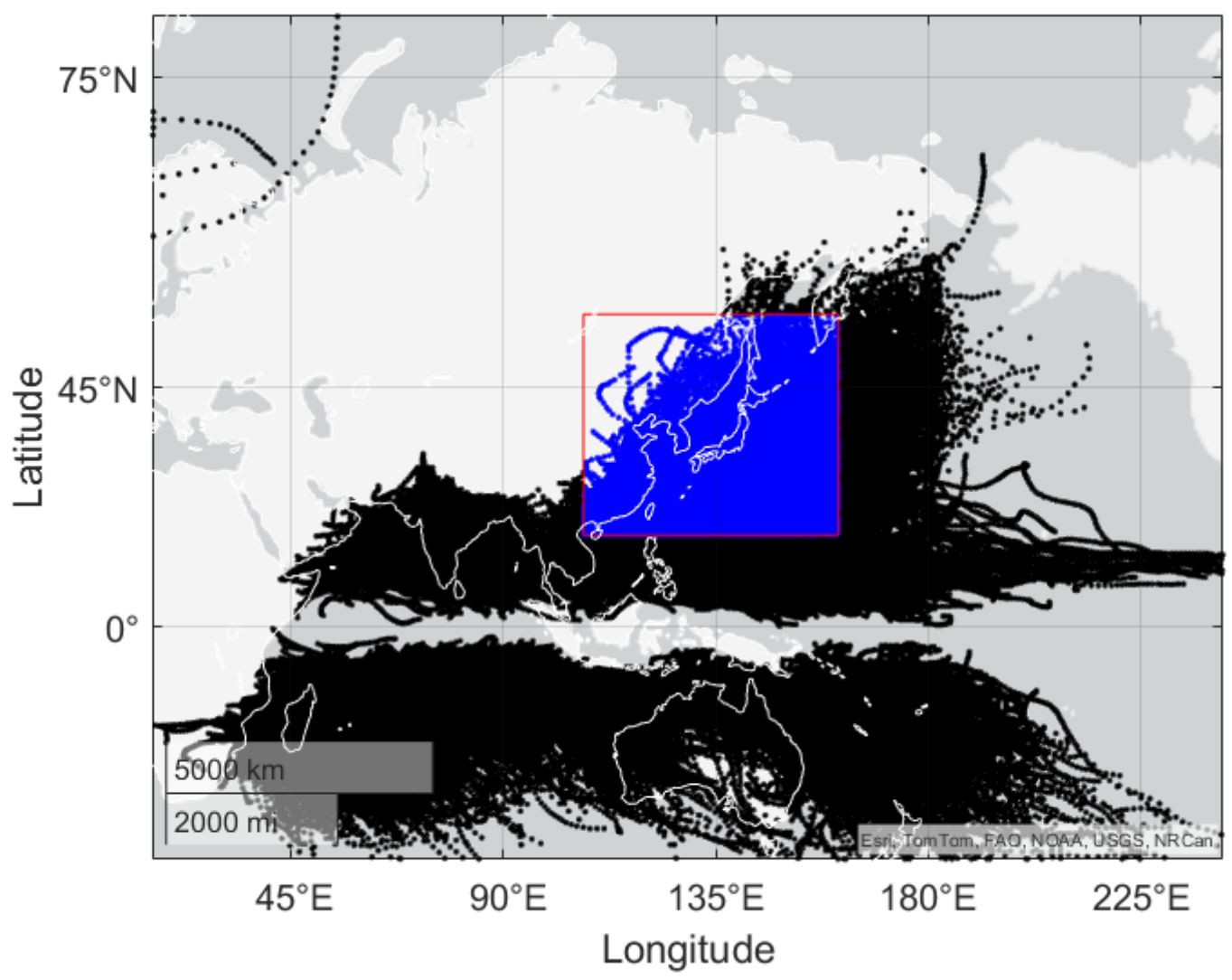

**Figure 13. Bounding box (red rectangle) and extracted storm events (in blue pattern).**

For each KHOA gauge site, wind speed and observed sea water level (O-sWL) data were batch downloaded. Rainfall data from KMA ground observations were also obtained and matched to the nearest KHOA gauge site. For each site, typhoon track data from IBTrACS passing within a 500 km radius were extracted. The daily maximum surge, daily maximum wind speed, and daily total rainfall corresponding to each typhoon event were then identified and paired based on matching calendar dates. It is noted that gauge records for some TC events are unavailable due to temporary instrument malfunctions. In the rest of this section, results of selected city sites of Busan, Wando and Tongyeong are present for discussion.

A correlation analysis was performed among the TC-induced surge, wind and rainfall. In particular, a lag–correlation analysis was conducted, and the results are shown in Figure 14. Here, the Kendall's tau is a rank correlation metric (in the range of [0,1]) typically used in copula analysis. The lag (in days) represents the temporal offset of one variable relative to the other. For example, in the surge–rainfall plot, a lag of +1 day indicates that the correlation is calculated between daily surge observations (days 1:n–1) and daily rainfall observations (days 2:n).

These results suggest that TC surges and winds intensify concurrently as a TC approaches and makes landfall, leading to the strongest correlation at zero lag. In contrast, rainfall often peaks earlier than the maximum surge and wind speeds, resulting in correlations that peak at a lag of –1 day. This temporal offset likely reflects the spatial structure and movement of the typhoon: heavy rainfall is typically concentrated in the leading rainbands of the storm, whereas maximum winds and storm surge occur closer to the time of closest approach or landfall.

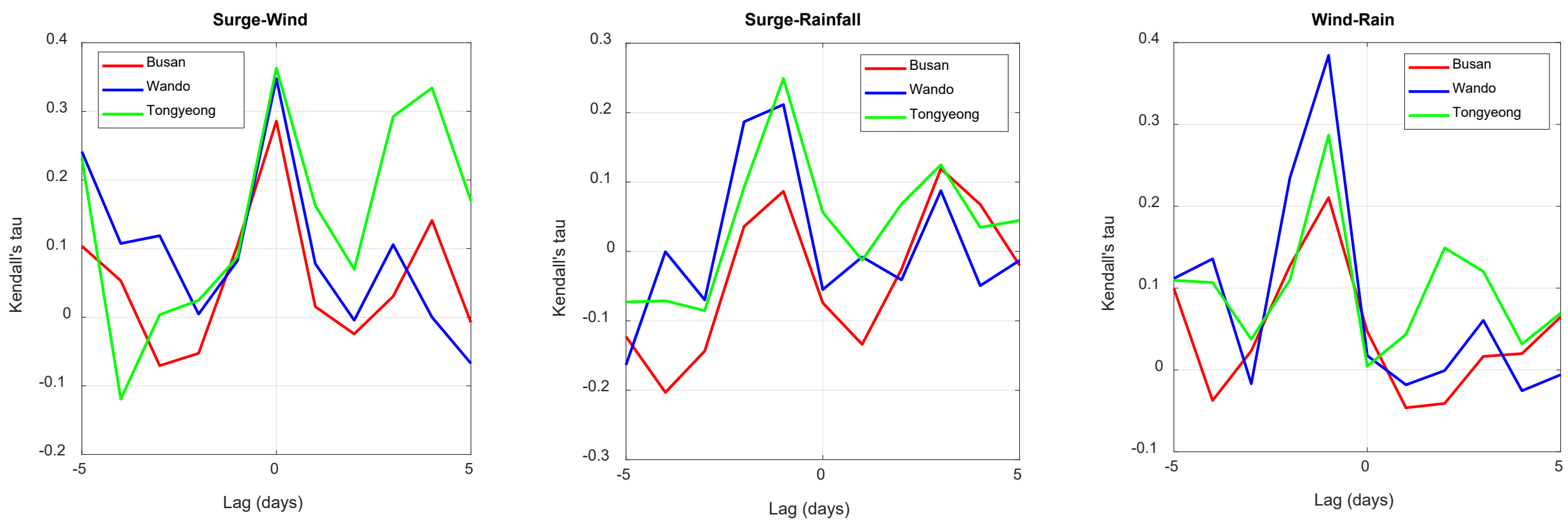

**Figure 14. TC-induced hazard correlation.**

To enable statistical analysis, it is important to remove temporal dependence in the time series and obtain statistically independent samples. For each TC event, missing records were first imputed using linear interpolation. Then, within each TC event period, the peak surge, peak wind speed, and peak daily rainfall were extracted to construct paired samples for subsequent analysis. Figure 15 presents the daily observations for all available TC events and highlights the extracted peak values for each variable used in the analysis.

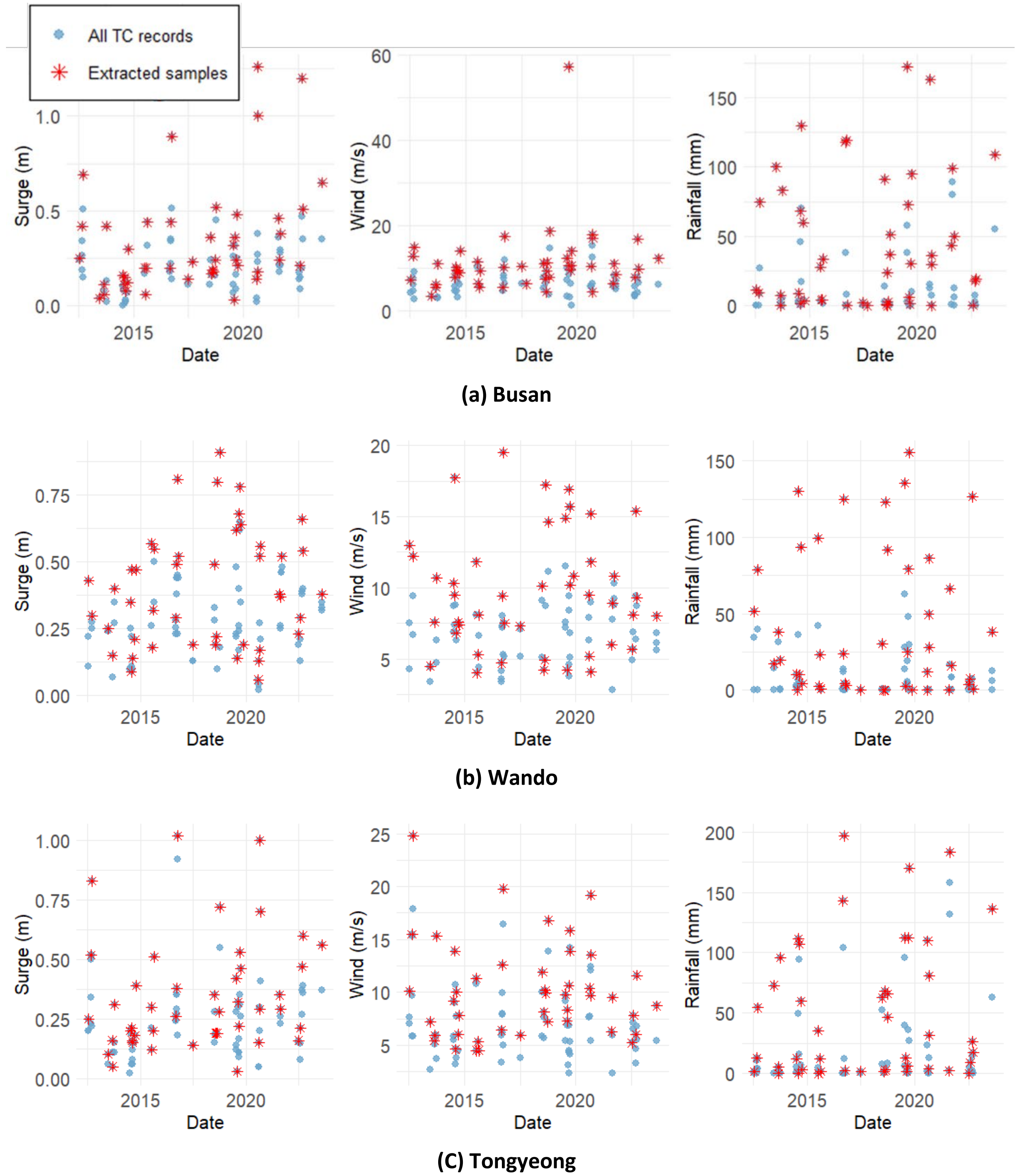

**Figure 15. Available TC events daily record and extracted samples.**

In this study, because TC-induced surge, wind, and rainfall exhibit positive correlations and the focus of this work is on investigating their compound effect, a three-dimensional Gumbel copula is employed to construct the joint distribution model.

The Gumbel copula is a member of the Archimedean copula family. Archimedean copulas have been widely applied in coastal and ocean engineering studies (Joe, 1997; Lin and Dong, 2019; Nelsen, 2007). The Gumbel copula is particularly suitable for modeling positive correlations and upper-tail dependence. Gamma distributions are employed as the marginal distributions for each variable, and the fitted probability density function (PDF) curves are shown in Figure 16. The fitted three-dimensional Gumbel copula based joint distribution is visualized by plotting pairwise PDF contours with the scatter plots of sample data, as shown in Figure 17.

It is noted that the marginal distributions of surge and wind exhibit a bell-shaped form, indicating the fitted Gamma distributions have shape parameters greater than one. In contrast, the rainfall marginal does not display a bell shape because of the large proportion of near-zero values, corresponding to a shape parameter smaller than one. These characteristics are also reflected in the pairwise joint PDF contour plots: the surge–wind contours appear more compact and elliptical, indicating a moderate or strong positive correlation, whereas the contours involving rainfall are relatively elongated and flattened, suggesting a relatively weak correlation.

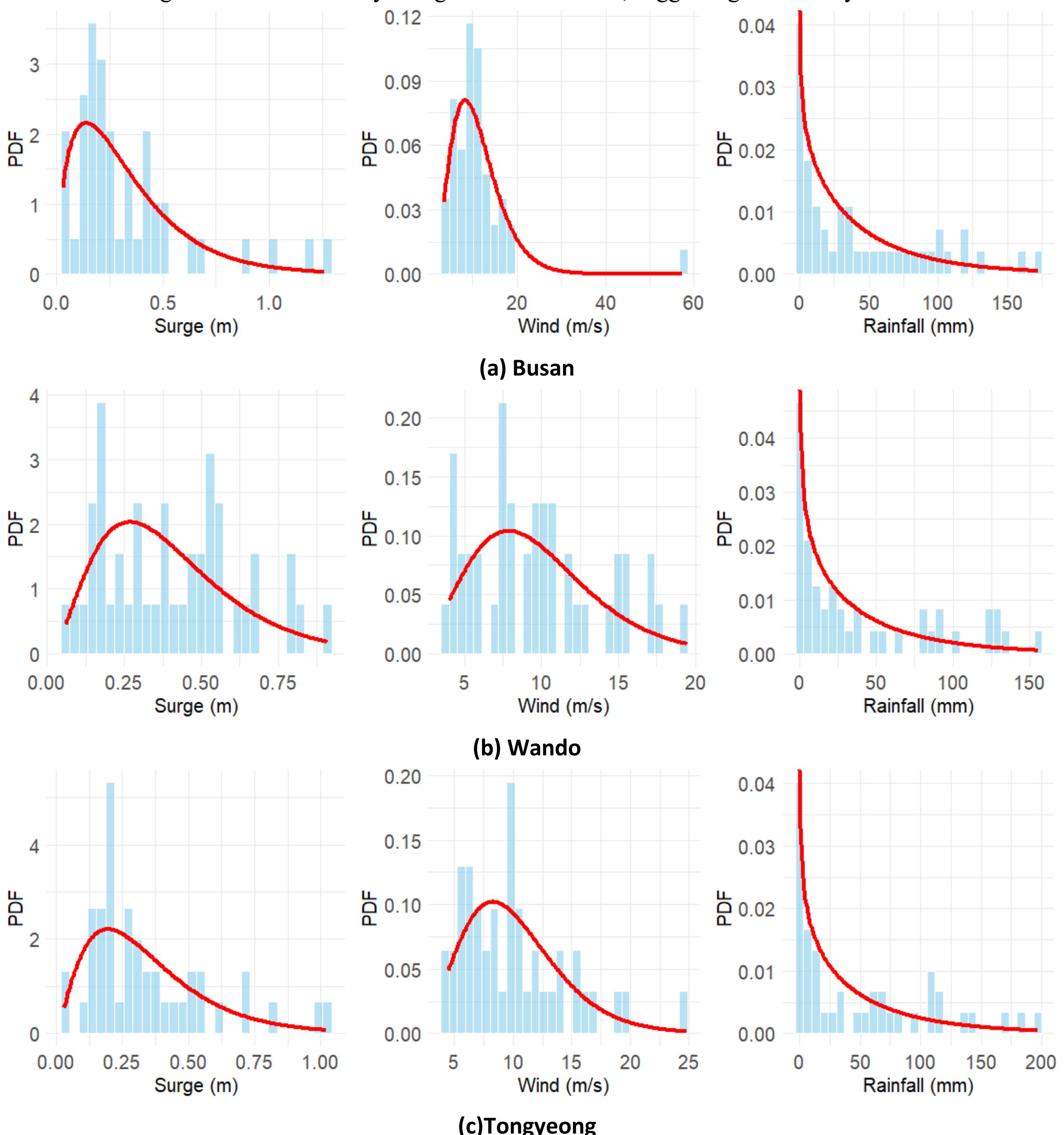

**Figure 16. Marginal analysis of each single hazard.**

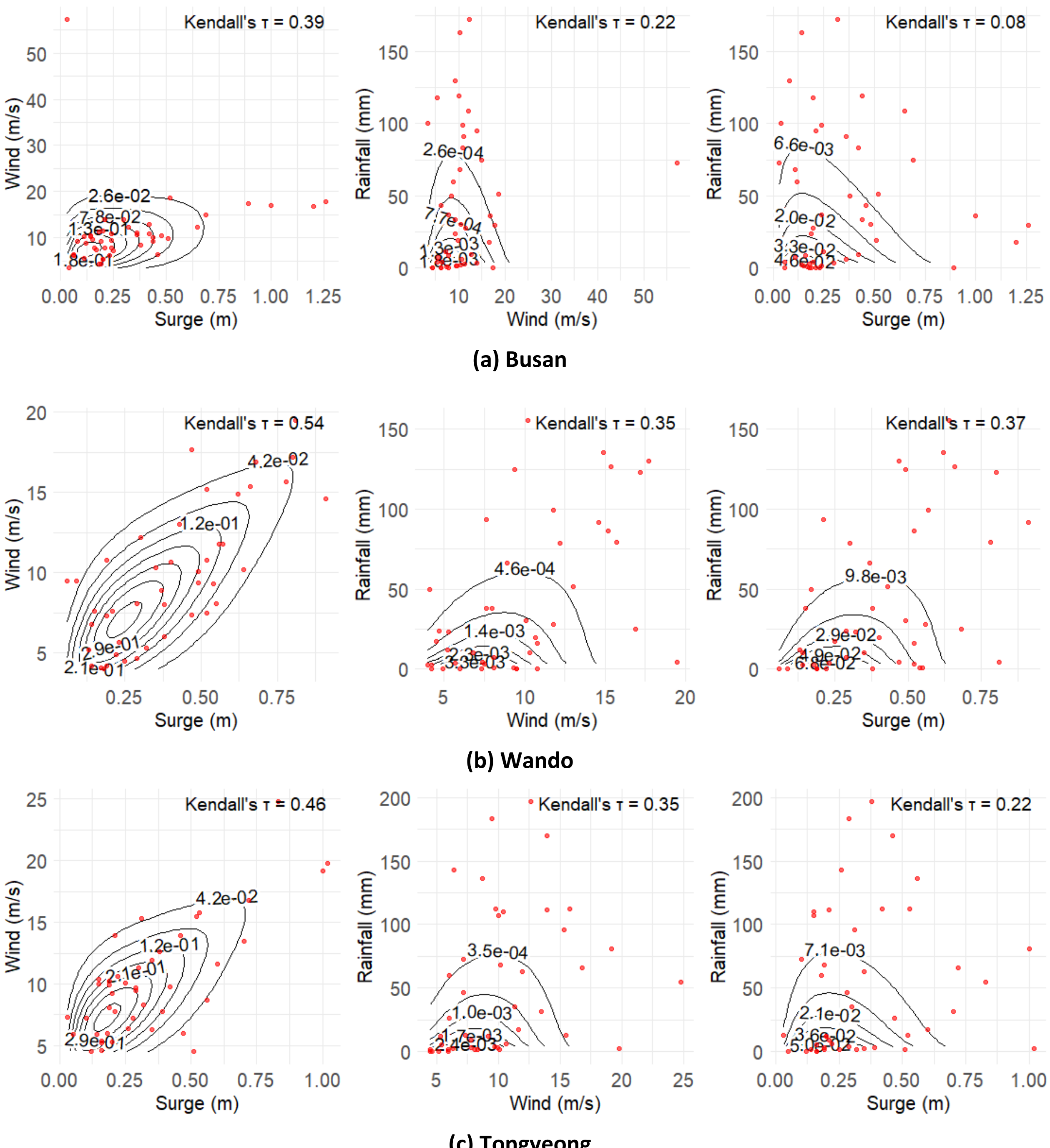

**Figure 17. Pair-wise PDF plot and scatter**

To evaluate the compound effect in TC-induced multi-hazard analysis, joint exceedance events are computed, tabulated, and compared with single-hazard exceedance events (i.e., exceedance events informed by marginal distribution of single-hazard). Following the trivariate analysis approach proposed by Wong et al. (2010), we estimate the exceedance probabilities of two types of joint events:

(1) the probability of exceeding any one of the thresholds for TC-induced surge, wind, or rainfall (i.e., the union joint event):

$$EP \cup = P(X_1 > x_1 \cup X_2 > x_2 \cup X_3 > x_3) = 1 - F(x_1, x_2, x_3) \tag{4}$$

(2) the probability of exceeding all of the thresholds for TC-induced surge, wind, or rainfall (i.e., the intersection joint event), the probability can be derived using DeMorgan's Rules as:

$$EP \cap = P(X_1 > x_1 \cap X_2 > x_2 \cap X_3 > x_3) = 1 - F(x_1, \infty, \infty) - F(\infty, x_2, \infty) - F(\infty, \infty, x_3) + F(x1, x2, \infty) + F(x1, \infty, x3) + F(\infty, x_2, x_3) - F(x_1, x_2, x_3) \quad (5)$$

The calculated exceedance probabilities are converted into return periods. The return periods of single-hazard events, derived from marginal analyses, and those of joint-hazard events derived from trivariate analysis are listed in Table 3. The threshold value for each type of TC-hazard (surge ($\eta$), windspeed ($w$) and rainfall ($p$)) are included for reference. The return period calculations account for the TC occurrence rate within the 500 km capture zone surrounding each gauge site. The results highlight the importance of incorporating compound effect of TC hazards into risk analysis.

Both the return periods for the union and intersection joint exceedance events reflect the positive correlations among surge, wind, and rainfall. For example, at Wando, a single-hazard event threshold with a 10-year return period based on univariate analysis may correspond to an intersection-type compound event with a return period of no more than 30 years. In contrast, if the correlations among hazards were ignored, the same intersection-type compound event compound event could appear to have a return period of approximately 1,000 years—demonstrating how correlation significantly influences joint exceedance estimates. It is also important to note the spatial variability among sites. In Busan, the return period of the intersection joint exceedance event is much longer than that at Wando or Tongyeong, which can be attributed to the relatively weaker correlations among surge, wind, and rainfall near the site.

**Table 3: Joint event return period table**

| Return period (years) of single hazard | Return period (years) of union joint event $EP \cup$ | Return period (years) of intersection joint event $EP \cap$ |
|---|---|---|
| Busan | | |
| 10<br>$(\eta = 0.97\ m,\ w = 24.0\ m/s,\ r = 171\ mm)$ | 3.75 | 116 |
| 100<br>$(\eta = 1.47\ m,\ w = 32.7\ m/s,\ r = 287\ mm)$ | 36.7 | 1220 |
| Wando | | |
| 10<br>$(\eta = 0.94\ m,\ w = 19.1\ m/s,\ r = 166\ mm)$ | 4.75 | 22.9 |
| 100<br>$(\eta = 1.33\ m,\ w = 25.2\ m/s,\ r = 284\ mm)$ | 46.8 | 231 |
| Tongyeong | | |
| 10<br>$(\eta = 0.88\ m,\ w = 19.6\ m/s,\ r = 199\ mm)$ | 4.25 | 27.9 |
| 100<br>$(\eta = 1.29\ m,\ w = 25.7\ m/s,\ r = 341\ mm)$ | 41.8 | 281 |

# 4 Conclusion and discussion

This study investigates the data sources, modeling approaches, and methodologies available for assessing typhoon-induced coastal hazards in South Korea. Typhoon-related observational and reanalysis data were collected from various agencies, with batch-download programs developed to interface with provider APIs. Visualizations were created to demonstrate the availability of historical storm information.

To address limitations in historical typhoon data, a data imputation method originally designed for the North Atlantic (Liu et al. 2024c) was applied to the WNP region. The availability and quality of observational O-sWL, rainfall, and river data were also evaluated. Key limitations were identified—for example, O-sWL data quality is relatively low before 2012, and publicly accessible typhoon records in KMA are only available from 2000 onward.

Leveraging the available data and the JPM approach, an SRR map was constructed for South Korea region, and a JPM analysis was performed for TC parameters in the region. Implementing the JPM for typhoon-induced surge hazard analysis requires efficient numerical modeling or the development of a surrogate prediction model. However, the limited availability of observational data or a comprehensive local storm model-derived dataset limits the ability to perform hazard severity prediction modeling. To address the constraints posed by this scarce historical data, systematic and well-designed hydrodynamic simulations are needed.

A trivariate TC-induced multi-hazards analysis was conducted using local gauge records, including surge, wind speed, and rainfall observations. Strong positive correlations were found among these variables, underscoring the importance of accounting for their interdependence and compound effects in TC hazard assessments in South Korea. Ignoring these correlations in single-hazard analyses may lead to underestimation of the hazard risk.

However, the limited length of available records—particularly O-sWL and wind speed data—may constrain the accuracy of statistical analyses, especially when estimating long return period events (e.g., those exceeding 100 years). Considering the limitations of historical data and the potential impacts of changing climate conditions in the region, developing efficient surrogate models for estimating typhoon responses (such as storm rainfall and storm surge) is essential for improving coastal hazard quantification in South Korea. To date, no publicly available high- or low-fidelity numerical modeling datasets have been identified to support the development of such surrogate models in this region.

## 5 Acknowledgment & Disclaimer

This work is supported by the Korea Atomic Energy Research Institute. KAERI Project Leads: Dr. Minkyu Kim and Dr. Eujeong Choi.

All opinions expressed in this work are those of the authors and do not necessarily reflect the policies or views of the research sponsor or any other organization.